\documentclass[10pt,conference]{IEEEtran}
\usepackage{graphicx}
\usepackage[utf8]{inputenc}
\usepackage{mathtools}

\let\subparagraph\relax
\usepackage{subfiles}
\usepackage[justification=centering,font=scriptsize,skip=1pt, belowskip=1pt,aboveskip=1pt]{caption}
\usepackage{color}
\usepackage{titlesec}
\usepackage{amsmath}
\usepackage{pgfplots}
\usepackage{xspace}

\let\OLDthebibliography\thebibliography
\renewcommand\thebibliography[1]{
  \OLDthebibliography{#1}
  \setlength{\parskip}{0pt}
  \setlength{\itemsep}{0pt plus 0.0ex}
}
\expandafter\def\expandafter\normalsize\expandafter
{%
    \normalsize
    \setlength\abovedisplayskip{2pt}
    \setlength\belowdisplayskip{0pt}
    \setlength\abovedisplayshortskip{0pt}
    \setlength\belowdisplayshortskip{0pt}
}

\titlespacing\section{0pt}{0pt plus 2pt minus 2pt}{2pt plus 2pt minus 2pt}
\titlespacing\subsection{0pt}{2pt plus 4pt minus 2pt}{2pt plus 2pt minus 2pt}
\titlespacing\subsubsection{0pt}{2pt plus 4pt minus 2pt}{2pt plus 2pt minus 2pt}

\usepackage{listings}
\usepackage{xcolor}

\definecolor{codegreen}{rgb}{0,0.6,0}
\definecolor{codegray}{rgb}{0.5,0.5,0.5}
\definecolor{codepurple}{rgb}{0.58,0,0.82}
\definecolor{backcolour}{rgb}{0.95,0.95,0.92}
\lstdefinestyle{mystyle}{
    backgroundcolor=\color{backcolour},   
    commentstyle=\color{codegreen},
    keywordstyle=\color{magenta},
    numberstyle=\tiny\color{codegray},
    stringstyle=\color{codepurple},
    basicstyle=\ttfamily\footnotesize,
    breakatwhitespace=false,         
    breaklines=true,                 
    captionpos=b,                    
    keepspaces=true,                 
    showspaces=false,                
    showstringspaces=false,
    showtabs=false,                  
    tabsize=2
}

\newcommand{\lightchain}{\textit{LightChain}\xspace}

\newcommand{\blockCapacity}{b\xspace}
\newcommand{\systemCapacity}{n\xspace}

\begin{document}

\title{A containerized proof-of-concept implementation of \lightchain system}
\author{
Yahya Hassanzadeh-Nazarabadi\IEEEauthorrefmark{1}\IEEEauthorrefmark{2}, 
Nazir Nayal\IEEEauthorrefmark{2},
Shadi Sameh Hamdan\IEEEauthorrefmark{2},
Öznur Özkasap\IEEEauthorrefmark{2},
and Alptekin Küpçü\IEEEauthorrefmark{2}\\
DapperLabs, Vancouver, Canada\IEEEauthorrefmark{1}\\
Department of Computer Engineering, Koç University, İstanbul, Turkey\IEEEauthorrefmark{2}\\
{\{yhassanzadeh13, nnayal17, shamdan17, oozkasap, akupcu\}}@ku.edu.tr}

\maketitle
\begin{abstract}
\lightchain is the first Distributed Hash Table (DHT)-based blockchain with a logarithmic asymptotic message and memory complexity. In this demo paper, we present the software architecture of our open-source implementation of  \lightchain, as well as a novel deployment scenario of the entire \lightchain system on a single machine aiming at results reproducibility.

\end{abstract}

\begin{IEEEkeywords}
Blockchain, Simulation, Docker, Distributed Hash Tables.
\end{IEEEkeywords}

\section{Introduction}
The existing blockchain architectures possess scalability problems concerning their asymptotic message and memory complexities \cite{croman2016scaling}. In particular, the well-known blockchain systems (e.g., Bitcoin \cite{nakamoto2008bitcoin}, Ethereum \cite{wood2014ethereum}, and Hyperledger \cite{cachin2016architecture}) operating with $\systemCapacity$ nodes impose a message complexity of O($\systemCapacity$) on disseminating a single transaction or block in the system. Moreover running a node in such systems requires a storage overhead of O($\blockCapacity$), where $\blockCapacity$ is the number of blocks in the system. Additionally, the existing blockchains' consensus protocols jeopardize the fairness and decentralization of the system by involving the nodes in the block generation decision-making based on their influence in the system, e.g., hashing power in Bitcoin's Proof-of-Work \cite{nakamoto2008bitcoin}, and amount of stakes in Ethereum's Proof-of-Stake \cite{wood2014ethereum}. 

To address the operational scalability issues of the blockchains, we have proposed \lightchain \cite{hassanzadeh2019lightchain}, which is the first fully decentralized Distributed Hash Table (DHT)-based blockchain architecture with $O(\log{\systemCapacity})$ message complexity for generating and disseminating a single transaction or block. \lightchain also provides a DHT-based replication technology for the blocks and transactions, which reduces the memory overhead of the nodes from $O(\blockCapacity)$ to $O(\frac{\blockCapacity}{\systemCapacity})$. At the consensus layer, \lightchain provides Proof-of-Validation, which is a fair and fully decentralized consensus protocol that provides each node a uniform chance of being involved in the block generation decision-making regardless of the node's influence in the system, e.g., its hashing power. 

In this demo paper, our contributions are two-fold. We first present the open-source implementation of a proof-of-concept for \lightchain in Java \cite{lightchain-container}, which represents a single \lightchain node. Running several instances of this implementation builds up a functional \lightchain system. Originally, each node is aimed to be deployed as a single container on the Google Cloud Platform's Kubernetes Engine \cite{hightower2017kubernetes}. However, reproducing the same deployment over the cloud is time-consuming and costly due to the pay-as-go policy of the cloud computing providers. Hence, as our second contribution, with the purpose of reproducibility, we present a containerized model of an entire system of \lightchain nodes that can be efficiently executed on a single machine. Our containerized implementation imitates the Kubernetes by orchestrating the nodes as well as their interactions based on the provided simulation configuration, plays the nodes against a simulation scenario, and collects and reports the performance of each node against predefined performance metrics of interest. Our containerized approach of \lightchain system can be used to locally orchestrate a simulation for any other distributed systems implementation on a single machine. We provide adaptation instructions in \cite{lightchain-container}.
\begin{figure}
\centering
\includegraphics[width=\linewidth]{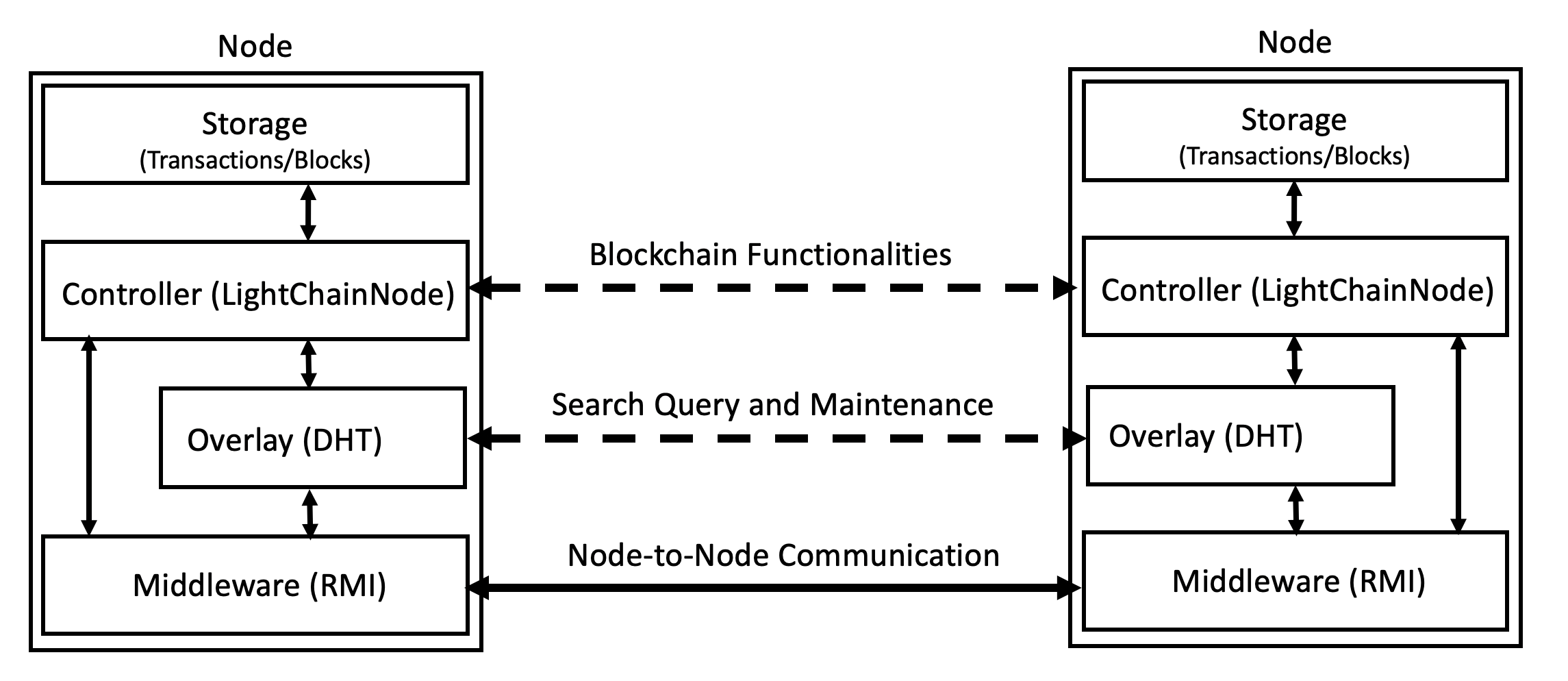}
\caption{Layered software architecture of \lightchain implementation. The solid horizontal arrow corresponds to a real path in the underlying network. The dashed horizontal arrows correspond to the virtual pairwise communication of the layers through the underlying network.}
\label{fig:layered-architecture}
\end{figure}

\section{Software Architecture}
Figure \ref{fig:layered-architecture} illustrates the layered software architecture of our \lightchain implementation. The interactions among the layers of a single node are shown by vertical double-sided arrows. Similarly, the interactions between levels of the same type on different nodes are depicted via horizontal arrows. Each node in our \lightchain implementation is made up of $4$ layers, which in a top-down approach are named as the Storage, Controller, Overlay, and Middleware. 

\textbf{Storage} layer represents an in-memory database of the transactions and blocks, and provides read and write functionalities for the lower Controller layer.

\textbf{Controller} implements the general behavior a \lightchain node. It observes the updates in the system through the overlay and conducts direct bidirectional blockchain-related communications with other nodes through the middleware layer, e.g., participating in the consensus. It also receives the transactions and blocks it is supposed to keep a replica through the overlay layer, maintains them in the storage layer, and answers queries of other nodes on those data objects.

\textbf{Overlay} implements a DHT node, which enables the controller to announce itself as well as its replicated data objects to other nodes in the system. The announcement is done by providing a unique identifier for the entity that it announces. The identifier of a controller is the hash value of its public key, and the identifier of the data objects is their hash value. Likewise, it enables the controllers of different nodes to search for each others' announcements in a fully decentralized manner. As the result of searching for a controller's public key, the overlay returns the address of the node that holds a controller with that public key. Similarly, as the result of searching for the identifier of a data object, the addresses of the nodes holding that data object in their storage layer are returned. 
It is worth noting that as the overlay is a DHT, both an announcement as well a search have a message complexity of $O(\log{\systemCapacity})$ \cite{aspnes2007skip}.  

\textbf{Middleware} provides direct inter-node communication in the underlying network. Each middleware instance is identified with a unique pair of IP address and port number. The middleware implements the Java Remote Method Invocation (RMI) \cite{pitt2001java} for the node. The middleware component of one node can directly communicate with the middleware instance of another node via Java RMI by solely knowing its address.  

\begin{figure}
\centering
\includegraphics[width=\linewidth]{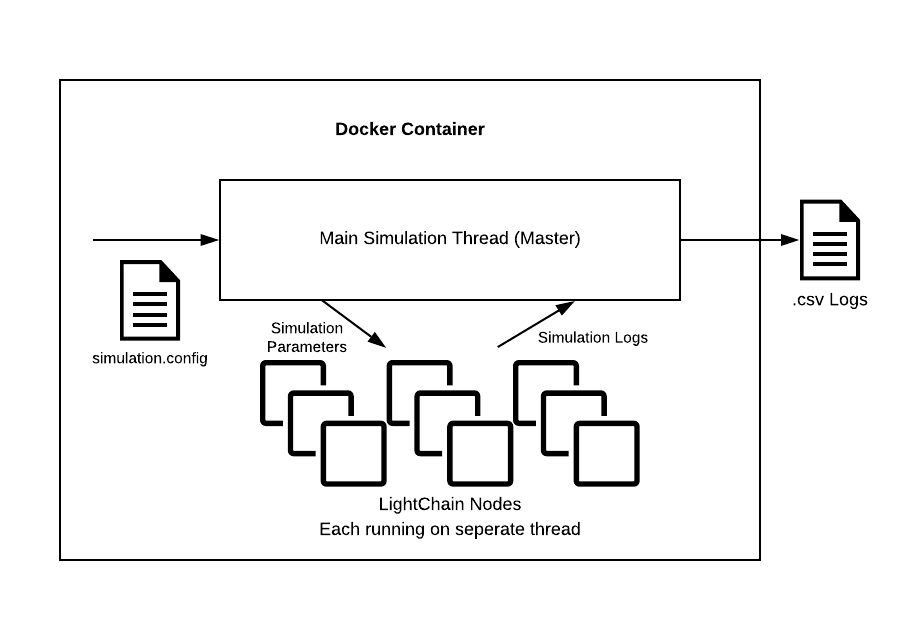}
\caption{Simulation pipeline of \lightchain system}
\label{fig:execution-pipeline}
\end{figure}

\section{Sample Demo Scenario}
Figure \ref{fig:execution-pipeline} shows a single containerized instance of the entire \lightchain system in our implementation. The entire system is encapsulated into a single docker container, which provides a virtualized operating system for it. Once the container starts, it executes the master thread. The master thread reads the \texttt{simulation.config} file, which lists up the simulation configuration for an entire \lightchain system. A sample configuration file is shown in Listing \ref{config}. This configuration describes a simulation scenario of $120$ nodes, where each node generates $1000$ transactions. The delay between every two consecutive transactions of a node is one second. A block in this simulation consists of a minimum of $100$ collected transactions. The initial balance of each node in the system is $20$ units, which is used to transfer funds to other nodes or pay the transaction processing fees. Also, $16\%$ of the nodes in this configuration are malicious nodes, that do not follow the \lightchain protocol. The second set of parameters in this listing are \lightchain specific parameters, which we skip for the sake of space, and refer the interested readers to \cite{hassanzadeh2019lightchain}.

Once the master thread loads the simulation configurations, it runs as many as the configured nodes with each node in a single thread, executes the simulation scenario for them and logs their interactions into an output \texttt{csv} file. Each log entry corresponds to a generated transaction or block and reports its memory, message, and operational time overheads. The simulation ends when each node generates its specified number of transactions, and each generated transaction appears into a block on the blockchain. The output \texttt{csv} file enables the user to perform arbitrary queries on the logs and extract performance metrics of the system, e.g., the average time it takes for a single node to generate a single transaction, the average time it takes for a single node to collect and generate a block of $100$ transactions, and the average block size of the system. To simulate the network latency between nodes, the master thread imposes a randomly generated latency for each pair of nodes, which follows our extracted distribution of pairwise latencies among the Google Compute Platform zones \cite{gcp-delay}.

\begin{lstlisting}[language=Java, caption= A sample \lightchain \texttt{simulation.config} file, label=config]
// Blockchain parameters
NODES = 120 // total nodes
TRANSACTIONS = 1000 // Total transactions per node
DELAY = 1 // delay between transactions of a node 
BLK_SIZE = 100 // size of a block
INIT_BALANCE = 20 // initial balance of a node
MALICIOUS = 0.16 // fraction of malicious nodes

// LightChain parameters
VALID_THR = 12 // signatures threshold
SIG_THR = 10 // validation threshold
VALID_FEE = 2  // validation fee
ROUTE_FEE = 1  // routing fee
REWARD = 3 // block generation reward
\end{lstlisting}

\bibliographystyle{IEEEtran}
\bibliography{references}
\end{document}